\documentclass{iopjournal}
\usepackage{graphicx} 
\usepackage{orcidlink}

\usepackage{iopams}
\usepackage{cite}

\def\c12{$^{12}\mathrm{C}(n,\alpha_0)$}
\def\o16{$^{16}\mathrm{O}(n,\alpha_0)$}
\def\po16{$^{16}\mathrm{O}(n,\alpha_{1,2,3})$}
\begin{document}

\articletype{Paper} 

\title{Differential cross sections for ${{^{12}\mathrm{C}(n,\alpha_{0})}}$, ${{^{16}\mathrm{O}(n,\alpha_{0})}}$ and ${{^{16}\mathrm{O}(n,\alpha_{1,2,3})}}$ between ${{E_n}}$ = 7.2 and 10 MeV with an active-target Time Projection Chamber}

\author{J. Bishop,$^{1, 2, *}$ \orcidlink{0000-0002-4701-8625}, C.E. Parker $^{2,a}$ \orcidlink{0000-0003-0362-6984}, R. Smith$^{3}$ \orcidlink{0000-0002-9671-8599}, Tz. Kokalova $^{1}$ \orcidlink{0000-0002-2035-3749}, G.V. Rogachev $^{2, 4, 5}$ \orcidlink{0000-0003-1120-6103}, C. Wheldon $^{1}$ \orcidlink{0000-0001-9137-6051}, S. Ahn $^{2, b}$ \orcidlink{0000-0001-8190-4914}, E. Koshchiy $^{2, c}$ \orcidlink{}, K. Brandenburg, $^{6}$ \orcidlink{0000-0002-9029-4019}, C.R. Brune $^{6}$ \orcidlink{0000-0003-3696-3311}, R.J. Charity $^{7}$ \orcidlink{0000-0003-3020-4998}, J. Derkin $^{6}$ \orcidlink{0000-0003-0014-5923}, N. Dronchi $^{8}$ \orcidlink{0000-0002-9429-0314}, G. Hamad $^{6}$ \orcidlink{0009-0009-8054-1483}, Y. Jones-Alberty $^{6, d}$ \orcidlink{0000-0002-0131-2204}, T.N. Massey $^{6}$ \orcidlink{0000-0002-5578-5357}, Z. Meisel $^{6, e}$ \orcidlink{0000-0002-8403-8879}, E.V. Ohstrom $^{8}$ \orcidlink{0000-0003-0913-6054}, S.N. Paneru $^{6}$ \orcidlink{0000-0001-7343-158X}, E.C. Pollacco $^{9}$ \orcidlink{0000-0003-4068-8682}, M. Saxena $^{6}$ \orcidlink{0000-0003-3430-7564}, N. Singh $^{6}$ \orcidlink{0000-0003-0898-8327}, L.G. Sobotka $^{7, 8, 10}$ \orcidlink{0000-0002-7883-7711}, D. Soltesz $^{6}$ \orcidlink{0000-0002-8243-3728}, S.K. Subedi $^{6}$ \orcidlink{0000-0001-5898-1989}, A.V. Voinov $^{6}$ \orcidlink{0000-0001-5918-7672} and J. Warren $^{6}$ \orcidlink{0000-0002-5294-3815}}

\affil{$^1$School of Physics and Astronomy, University of Birmingham, B15 2TT, UK}

\affil{$^2$Cyclotron Institute, Texas A\&M University, College Station, TX 77843, USA}

\affil{$^3$School of Engineering and Built Environment, Sheffield Hallam University, Sheffield, S1 1WB, UK}

\affil{$^4$Department of Physics \& Astronomy, Texas A\&M University, College Station, TX 77843, USA}

\affil{$^5$Nuclear Solutions Institute, Texas A\&M University, College Station, TX 77843, USA}

\affil{$^6$Edwards Accelerator Laboratory, Department of Physics \& Astronomy, Ohio University, Athens, OH 45701, USA}

\affil{$^7$Department of Chemistry, Washington University, St. Louis, MO 63130, USA}

\affil{$^8$Department of Physics, Washington University, St. Louis, MO 63130, USA}

\affil{$^9$IRFU, CEA, Universit\'e Paris-Saclay, Gif-Sur-Yvette, France}

\affil{$^{10}$McDonnell Center for the Space Sciences, Washington University, St. Louis, MO 63130, USA}

\affil{a Present Address: Edwards Accelerator Laboratory, Department of Physics \& Astronomy, Ohio University, Athens, OH 45701, USA}

\affil{b Present Address: Center for Exotic Nuclear Studies, Institute for Basic Science, 55, Expo-ro, Yuseong-gu, Daejeon, 34126, Republic of Korea}

\affil{c Deceased}

\affil{d Present Address: The Johns Hopkins University Applied Physics Laboratory (JHU/APL), Laurel, MD 20723, USA}

\affil{e Present Address: Department of Engineering Physics, Air Force Institute of Technology}

\affil{$^*$Author to whom any correspondence should be addressed.}

\email{j.bishop.2@bham.ac.uk}


\begin{abstract}

Data for the \c12, \o16, and \po16 differential cross sections are important for several different areas of nuclear physics such as understanding neutron transmutation in nuclear reactors.

The TexAT Time Projection Chamber was used to measure the differential and angle-integrated cross sections in active-target mode. The chamber was filled with CO$_2$ gas and used a quasi-monoenergetic neutron beam from the $d(d,n)$ reaction at Edwards Accelerator Lab at Ohio University.

A comparison between our current and previous results at overlapping energies and angles showed good agreement in angular dependence and absolute cross section. A broader angular coverage than previous results demonstrated that the integrated cross section for the \po16 reaction deviates from ENDFVIII.0 evaluations.

This first instance of neutron-induced measurements with an active-target Time Projection Chamber demonstrates the use of this method for high-quality differential cross section data across a broad angular range, generating good statistics with a relatively low-intensity beam.
\end{abstract}

\section{\label{sec:Introduction}Introduction}
A high-precision understanding of $(n,\alpha)$ cross sections is essential for many areas of nuclear physics. Of particular importance is the $^{16}\mathrm{O}(n,\alpha_0)$, which is a crucial ingredient in understanding the neutron multiplication factor in nuclear reactors, $k_{eff}$. Additionally, this reaction accounts for 25\% of total helium production, affecting the performance of fuel pins and clads \cite{NEA}. Discrepancies between different measurements mean that, while the energy dependence of this cross section is reasonably understood, the overall normalisation was not known to better than 30\% due to issues with historical data. It is for this reason that a measurement of the total cross section for this reaction is at the top of the UK Nuclear Data High Priority request list \cite{HPRL,HPRLweb} which motivated this work. Recent studies from LANL \cite{Lee} have provided data for differential cross sections at a small number of angles; however, the full angular range has not been measured.  \par
Additionally, the \c12 reaction is important for neutron spectroscopy and understanding the delivered dose in the body for fast neutron therapy. Using an active-target approach with diamond detectors, the \c12 total and differential cross sections have been previously measured across a broad range of neutron energies, with the total cross section being well understood \cite{12CnaLANSCE,Wantz}. Later works have also measured the $^{12}\mathrm{C}(n,n+3\alpha)$ cross section which has contributions from $^{12}\mathrm{C}(n,n')^{12}\mathrm{C}^{\star}$, $^{12}\mathrm{C}(n,\alpha)^{9}\mathrm{Be}^{\star}$ as well as $^{12}\mathrm{C}(n,^{5}\mathrm{He})^{8}\mathrm{Be}$ \cite{Liu}. This active-target approach is particularly efficient and the neutron-induced reactions on oxygen cannot be measured with a solid active-target approach in the same way. However, it is possible to use a gas target containing both carbon and oxygen to measure these reactions simultaneously. This provides additional cross normalisation and validation. With limited neutron fluence, gaining sufficient statistics for differential cross sections typically requires extremely specialised facilities or extended beam time.\par
To overcome these difficulties, this work took advantage of a Time Projection Chamber (TPC) which has both high target thickness and 4$\pi$ solid-angle coverage to measure the neutron-induced reactions on both C and O from carbon dioxide (CO$_2$) counting gas. Using a pseudo-monoenergetic neutron source from the Edwards Accelerator Laboratory at Ohio University \cite{EAL} from a $d(d,n)$ reaction with a tunable deuteron energy, five different reaction cross sections from neutron-induced reactions were measured simultaneously and differentiated inside the TPC. The first two were the $^{12}\mathrm{C}(n,n_{2})3\alpha$ and $^{12}\mathrm{C}(n,\alpha_{1,2})^{9}\mathrm{Be}$ - i.e. the inelastic scattering of carbon-12 to the second excited state (the Hoyle state), and the $(n,\alpha')$ populating unbound states in $^{9}\mathrm{Be}$. Both of these reactions yield a final state of 3 $\alpha$-particles and a neutron and have been previously published \cite{JBNature}. In addition to these 3$\alpha$+n final-state cross sections, the important two-body final state reactions $^{12}\mathrm{C}(n,\alpha_{0})$, $^{16}\mathrm{O}(n,\alpha_{0})$ and $^{16}\mathrm{O}(n,\alpha_{1,2,3})$ (where $\alpha_{1,2,3}$ indicates $\alpha_1$, $\alpha_2$ and $\alpha_3$ which cannot be separated due to the similar excitation energies in $^{13}$C) were measured between neutron energies of 7.2 and 10 MeV. The first of these, $^{12}\mathrm{C}(n,\alpha_{0})$, is used to benchmark our methods against the well known LANSCE results \cite{12CnaLANSCE} and therefore constitutes an excellent benchmark to validate any $^{16}\mathrm{O}(n,\alpha)$ cross section results. 
\begin{figure*}
    \centering
    \includegraphics[width=0.95\textwidth]{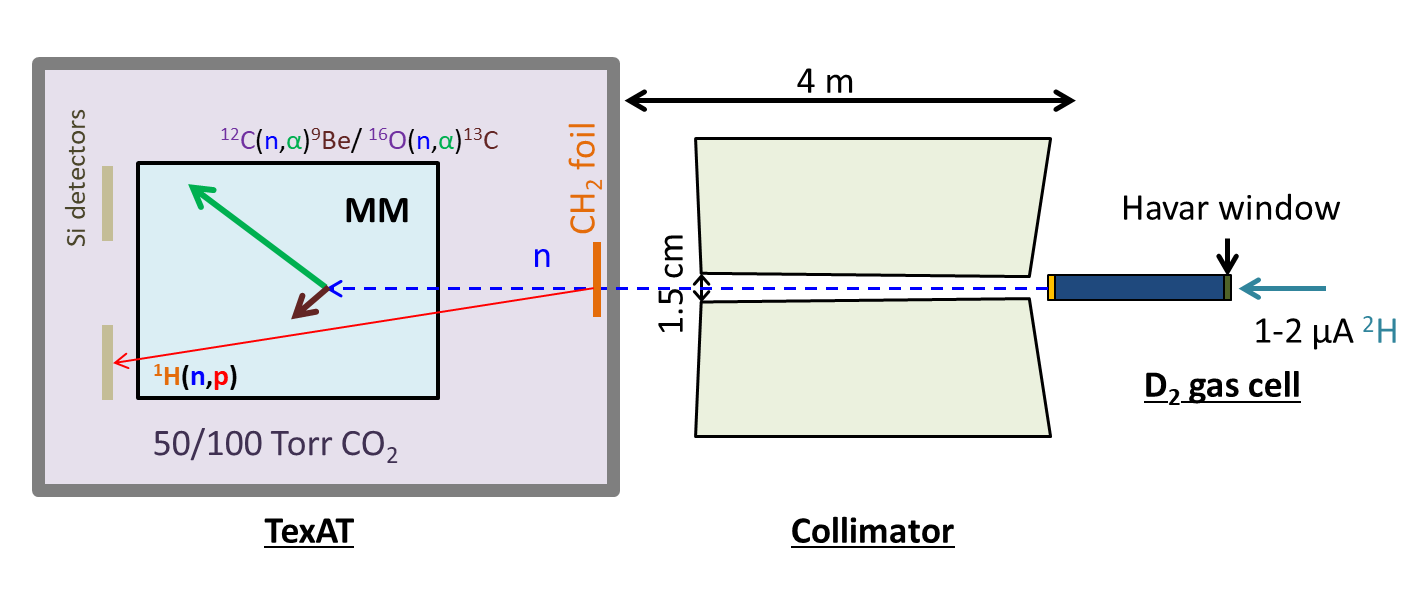}
    \caption{Schematic overview showing the key features of the setup. The neutron beam was generated by a deuteron beam incident on a deuterium gas cell. A collimator system reduces the size of the beam at 4 meters to 1.5-cm diameter. The neutrons are then incident upon the TexAT TPC where (n,$\alpha$) reactions are measured from the CO$_2$ gas inside the active area of the Micromegas (MM). In addition, neutron elastic scattering off a CH$_2$ foil at the entrance to the detector was used for normalisation by measuring the protons in Si detectors in the forward direction.}
    \label{fig:setup}
\end{figure*}
\section{Experimental Setup}
To study neutron-induced reactions, a direct-current deuteron beam of 1-2~$\mu$A from the 4.5-MV tandem accelerator at Edwards Accelerator Laboratory, Ohio University was provided. This high-intensity deuteron beam was incident on a deuterium-filled gas cell with a total length of 7.98 cm at a pressure of 760 Torr, isolated from the beamline by a 2.5-$\mu$m-thick Havar foil. To stop the unreacted beam, a gold foil was placed at the end of the gas cell. \par
The forward-focused neutrons from the $d(d,n)$ reaction were collimated by a multi-material system placed in the 30-m time-of-flight tunnel. A 0.75-cm-diameter opening aperture ensured that the neutron beam-spot size had a diameter of 0.75 cm at a distance of 2 m from the gas cell, increasing to 1.5 cm at the exit of the collimator. In order to provide neutron energies between 7.2-10~MeV (with widths ranging from 0.35-0.25 MeV), the incident deuteron energy was varied from 4.25 to 7~MeV. Previous studies using the same dataset have provided information on the energy verification using a NE-213 detector placed 30 m from the gas cell \cite{JBNature}. The generated neutron energy spectrum is non-Gaussian due to being dominated by the energy loss through the deuterium gas cell with the exact energy spectra available with the accompanying dataset \cite{edata}and is between 2-8\% depending on the neutron energy. The typical intensity of the monoenergetic neutron beam was 5000 neutrons per second after the collimator. The low neutron beam intensity therefore necessitates the use of a detector system with high efficiency and target thickness. This was achieved using the TexAT Time Projection Chamber (TPC) \cite{TexAT}, which was placed at a distance of 4 m from the gas cell and was filled with either 50 or 100 Torr of CO$_2$ gas (50 Torr for $E_n$ = 8.36 - 10 MeV and 100 Torr for $E_n$ = 7.25 - 8.15 MeV) where the higher pressure was chosen to optimise statistics. Interactions in the gas produce charged-particles which lose energy in the gas via creation of electron-ion pairs. These electrons are drifted by the field cage which has an electric field of 69 V/cm corresponding to an electron drift velocity of 0.75 cm/$\mu$s. The signals from the drifted electrons are amplified in two stages, first by a Thick GEM \cite{GEM}(thickness 1.5 mm) and then by Micromegas \cite{MM1,MM2} (mesh height 128 $\mu$m). This corresponded to an overall gas gain of around 5000. The signal is then read out by the 1024 channels of the Micromegas plane (size 245 x 224 mm$^2$). The trigger condition required the signal in only one pad to be above threshold. In addition to the Micromegas, four 625-$\mu$m-thick Si detectors were placed 50 cm from the beam entrance. These were used for normalisation where elastically scattered protons from a 30- (or 60- for some energies)~$\mu$m-thick CH$_2$ foil placed at the beam entrance were measured. The cross section for the $^{1}\mathrm{H}(n,p)$ is well-known and provided the overall flux of neutrons for each energy. These Si detectors were also placed in the trigger where a silicon hit or the Micromegas triggered an event.
Figure~\ref{fig:setup} shows an overview of the setup.

\section{Channel separation}
Inside of the TexAT TPC, the 3D tracks of the resultant particles were reconstructed using the RANSChiSM method \cite{RANSChiSM}. Given the trigger multiplicity required only one pad to fire, a large portion of events originated from $^{12}\mathrm{C}(n,n_0)$ and $^{16}\mathrm{O}(n,n_0)$ elastic scattering. The recoil $^{12}\mathrm{C}$ or $^{16}\mathrm{O}$ has a relatively small energy and therefore a small range. These events were removed by cutting events where the length of any tracks was shorter than the maximum allowed range for elastic scattering events. For all scattering angles and energies, the $(n,\alpha)$ events of interest $-$ $^{12}\mathrm{C}(n,\alpha)$, $^{16}\mathrm{O}(n,\alpha_0)$ and $^{16}\mathrm{O}(n,\alpha_{1,2,3})$ $-$ had longer tracks than the elastic scattering limit but for the lower neutron beam energies, the separation between the channels became problematic. \par
The RANSChiSM technique fits events with two tracks, one for the light particle ($^{4}\mathrm{He}$) and one for the heavy particle ($^{9}\mathrm{Be}/^{13}\mathrm{C}$). To parameterise this fit, three locations are chosen corresponding to: the interaction vertex (which was limited to lie within the known region where the neutron beam was incident), the end point of the light track, and the end point of the heavy track. The RANSChiSM technique chooses these points by randomly sampling the available point cloud and the best fit is selected by choosing the fit lines that minimise the chi-squared. Compared to previous implementations of RANSChiSM, an additional term is included to ensure conservation of momentum transverse to the beam direction among the final particles with the penalty to the total $\chi^2$ scaling as $(1-|\vec{a}.\vec{b}|)$, where $\vec{a}$ and $\vec{b}$ are the vectors of the light and heavy particle along the xy direction. Once the optimum fit has been achieved, the kinematics of the decay are characterised by evaluating four variables. These variables are: $\theta_{L}$, the light-recoil lab angle relative to the beam, $\theta_{H}$, the heavy-recoil angle relative to the beam, $R_{L}$, the light-recoil range in the TPC active region and $R_{H}$, the heavy-recoil range in the TPC active region. For a large number of events, the $\alpha$-particle had sufficient energy from the decay to escape the active volume of the TPC. Therefore, the range of the light fragment, $R_{L}$, is often a lower limit of the true range and is not a meaningful cut. Two different selection methods are then made to differentiate between the three channels where the channel with the smallest chi-squared is initially taken. As the last 5~mm of the tracks for both $^{12}$C and $^{16}$O are below threshold, a range extending correction to the measured $R_H$ was applied. \par
\begin{figure}
    \centering{\includegraphics[width=0.5\textwidth]{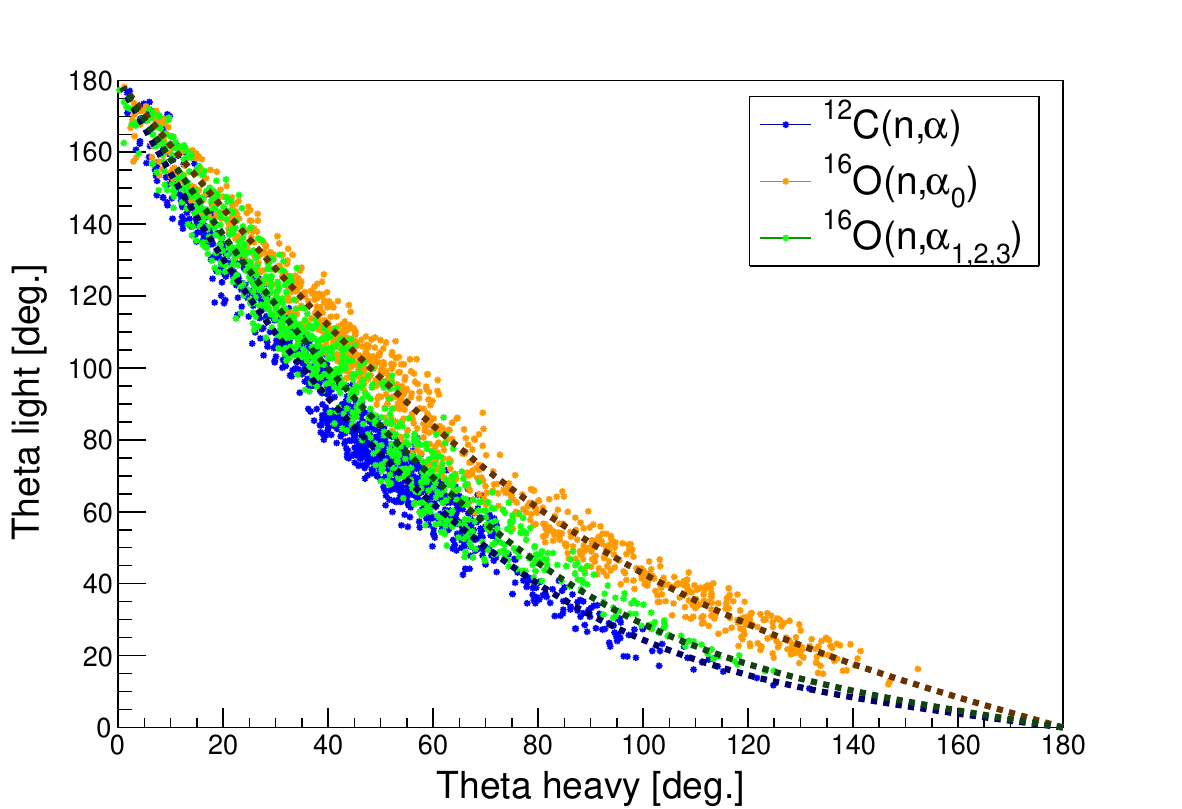}\includegraphics[width=0.5\textwidth]{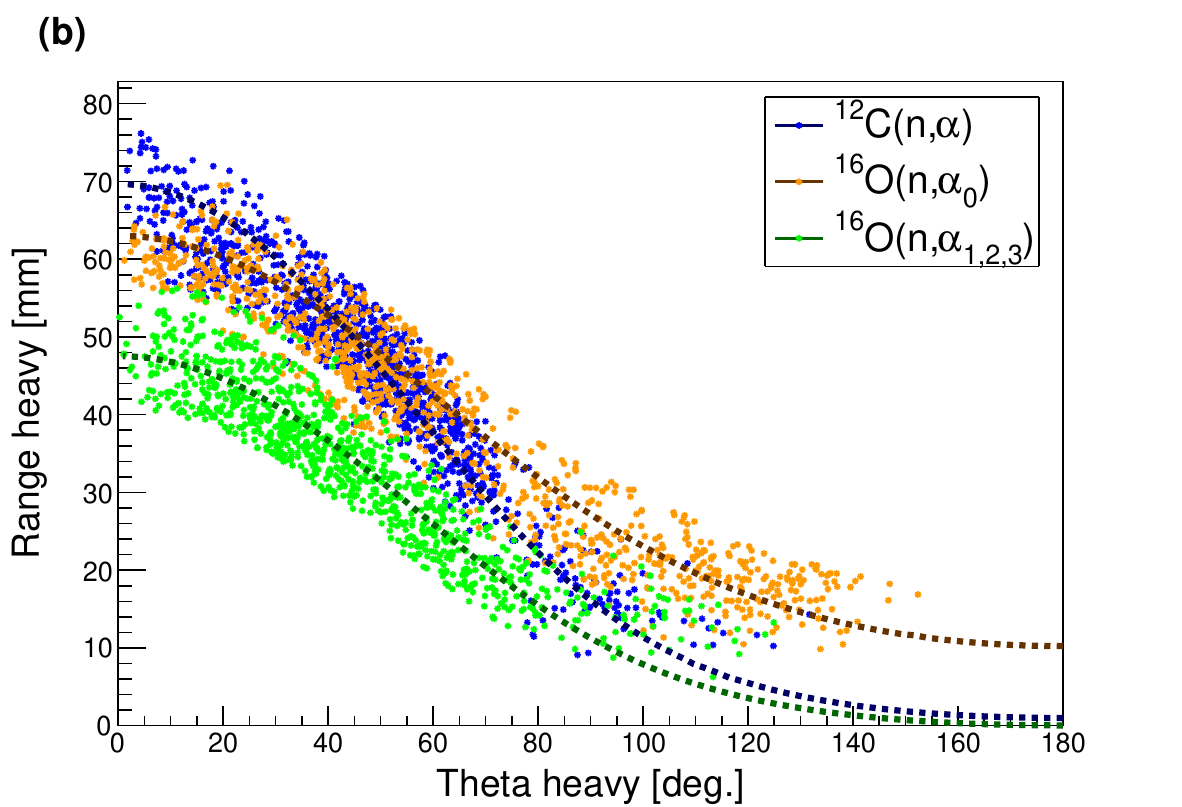}}

    \caption{Example final separation of events by variables for $E_n$ = 9.25 MeV for heavy lab angle vs light lab angle (left) and heavy range against heavy lab angle (right) with events classified as \c12 denoted by blue points, \o16 events by orange points, and \po16 by green points. The kinematic predictions for each are also overlaid with the same colours in a dotted line of a darker hue.}
    \label{fig:Ed6p25}
\end{figure}

The first selection is based on the relationship between $\theta_{L}$ and $\theta_{H}$. All three channels demonstrate very similar relationships between these quantities, with the left plot of Fig.~\ref{fig:Ed6p25} showing the experimental data overlaid with loci for the three different channels. The angle-angle map can distinguish $^{16}\mathrm{O}(n,\alpha)$ from the other two channels, but the separation is not unambiguous at all angles and another channel selection criteria is required. \par
The second selection is based on the relationship between $\theta_H$ and $R_{H}$, which is more reliable than utilising the light-particle track as the heavy particle cannot escape the active region. The right plot of Fig.~\ref{fig:Ed6p25} shows the data overlaid with the loci for each of the different channels and a clear separation can be seen between $^{12}\mathrm{C}(n,\alpha_0)/^{16}\mathrm{O}(n,\alpha_0)$ and \po16 with some difficulties arising at small $\theta_H$. Combining the two selections therefore allows for separation of the three channels.\par
On an event-by-event basis, evaluating the $\chi^2$ for each of these two selections against the expected kinematics allows for a probabilistic measurement of the cross sections with uncertainty propagation. The motivation for this is that the channel separation becomes poor at certain angles and some events from a dominant channel mischaracterised as a weaker channel can lead to incorrect differential cross sections.\par
For the channel $k$ for event $i$, the chi-squared $\chi_{ik}^2$ is then converted to a classification probability, $p_{ik}$. These probability values are then normalised to relative probabilities \cite{Bayes}, $P_{ik}$, for each channel so that:
\begin{eqnarray}
\Sigma_{k} P_{ik} = 1.
\end{eqnarray}
This allows for the formulation of a multinomial distribution for the three channels event by event where the yield is determined to be:
\begin{eqnarray}
Y_k = \Sigma_i P_{ik},
\end{eqnarray}
and the variance is given as:
\begin{eqnarray}
\sigma^2_k = \Sigma_i P_{ik}(1-P_{ik}).
\end{eqnarray}
As an example, if each channel were as likely as the others (p=$\frac{1}{3}$) then the yield for each channel would increase by $\frac{1}{3}$ but the error bars would increase by $\frac{\sqrt{2}}{3}$, signifying the degree of uncertainty. With $p=0.9$, the yield for the most likely channel increases by 0.9 with the error bars increasing by 0.3. In this approach, the differential cross section for each channel can be evaluated by counting the yield in the centre of mass (c.o.m) for $\theta_{c.o.m}$ with the different correct conversions from $\theta_{lab}$ for the three channels.
\par

\section{Differential cross sections}
Following the conversion from lab to centre of mass angle, the c.o.m. differential cross sections were calculated for the three channels for the different neutron energies. As the interaction vertex is hard to determine for decays closely aligned to the beam axis, the angular range for reliable differential cross sections is restricted. The covered range for good-track events is therefore limited to 40-160$^{\circ}$ in the centre of mass by this constraint. Additionally, as with the $^{12}\mathrm{C}(n,\alpha_0)$ and $^{16}\mathrm{O}(n,\alpha_1)$ reactions which have quite large negative Q-values, the heavy fragment range for decays with a heavy-fragment laboratory angle above 90$^{\circ}$ is very small ($<$ 10 mm) such that these data may not be sufficiently reliable due to track fitting difficulties. For each energy, the safe angular range for each decay path was evaluated using the two above constraints for each decay path, neutron energy, and TPC gas pressure. \par
The energy error bars plotted throughout the paper correspond to the total width of our incident neutron energy spectrum due to the non-Gaussian beam energy profile. The differential cross section errors represent the channel separation statistical uncertainty and Poisson statistics. The normalisation uncertainty from the counting statistics of $^{1}$H$(n,p)$ events incorporates a systematic uncertainty that is represented in the full data files accompanying this paper \cite{edata}.
\subsection{$^{12}\mathrm{C}(n,\alpha_0)$}

\begin{figure}[ht!]
    \centering{\includegraphics[width=\textwidth]{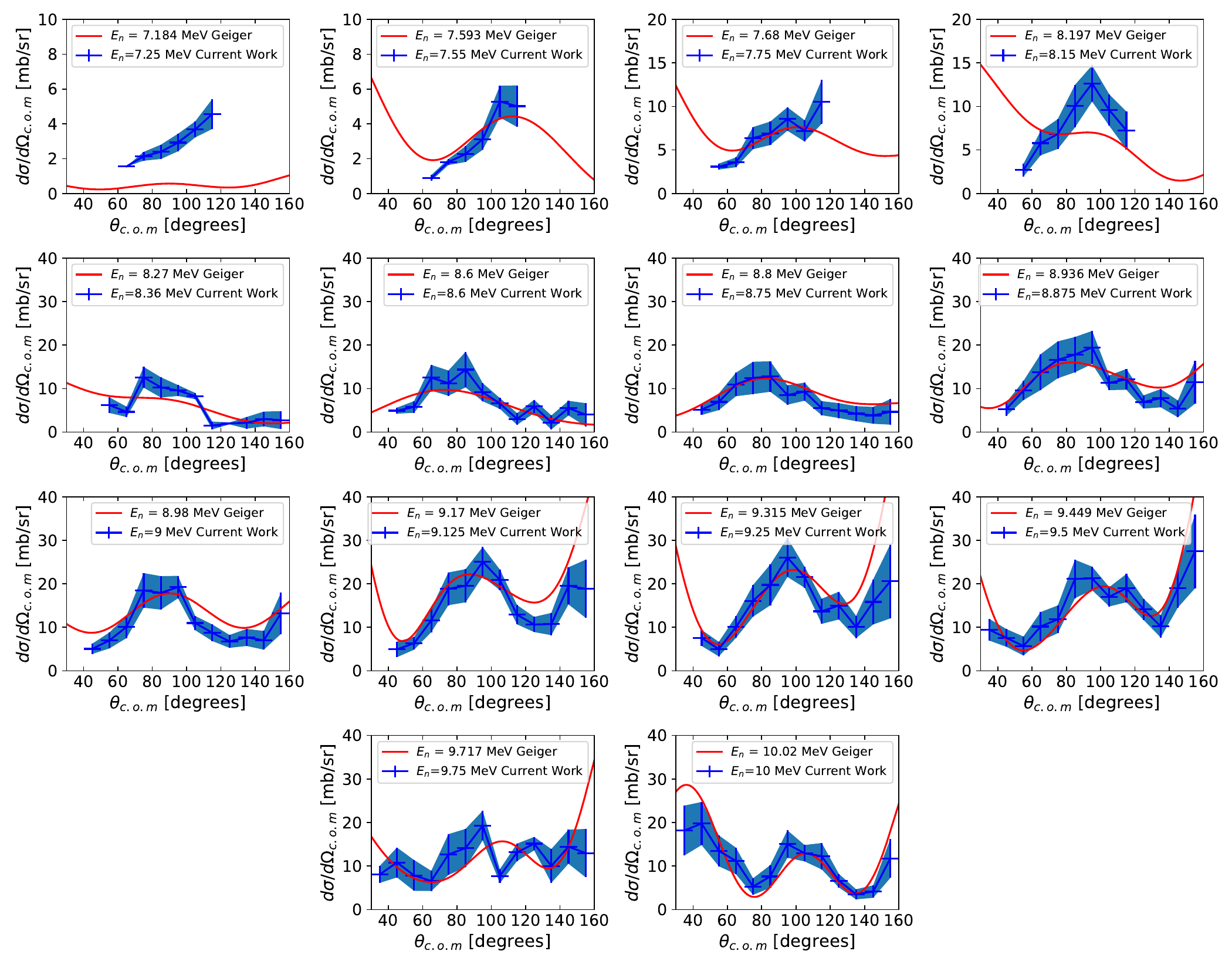}}
    \caption{Centre-of-mass differential cross sections for the \c12 reaction at different incident neutron energies, in comparison to the Legendre coefficients of Geiger \cite{Geiger}, normalised to reproduce the total cross sections measured by Kuvin \cite{12CnaLANSCE}.}
    \label{fig:diff12C}
\end{figure}

\begin{figure}[ht!]
    \centering{\includegraphics[width=0.5\textwidth]{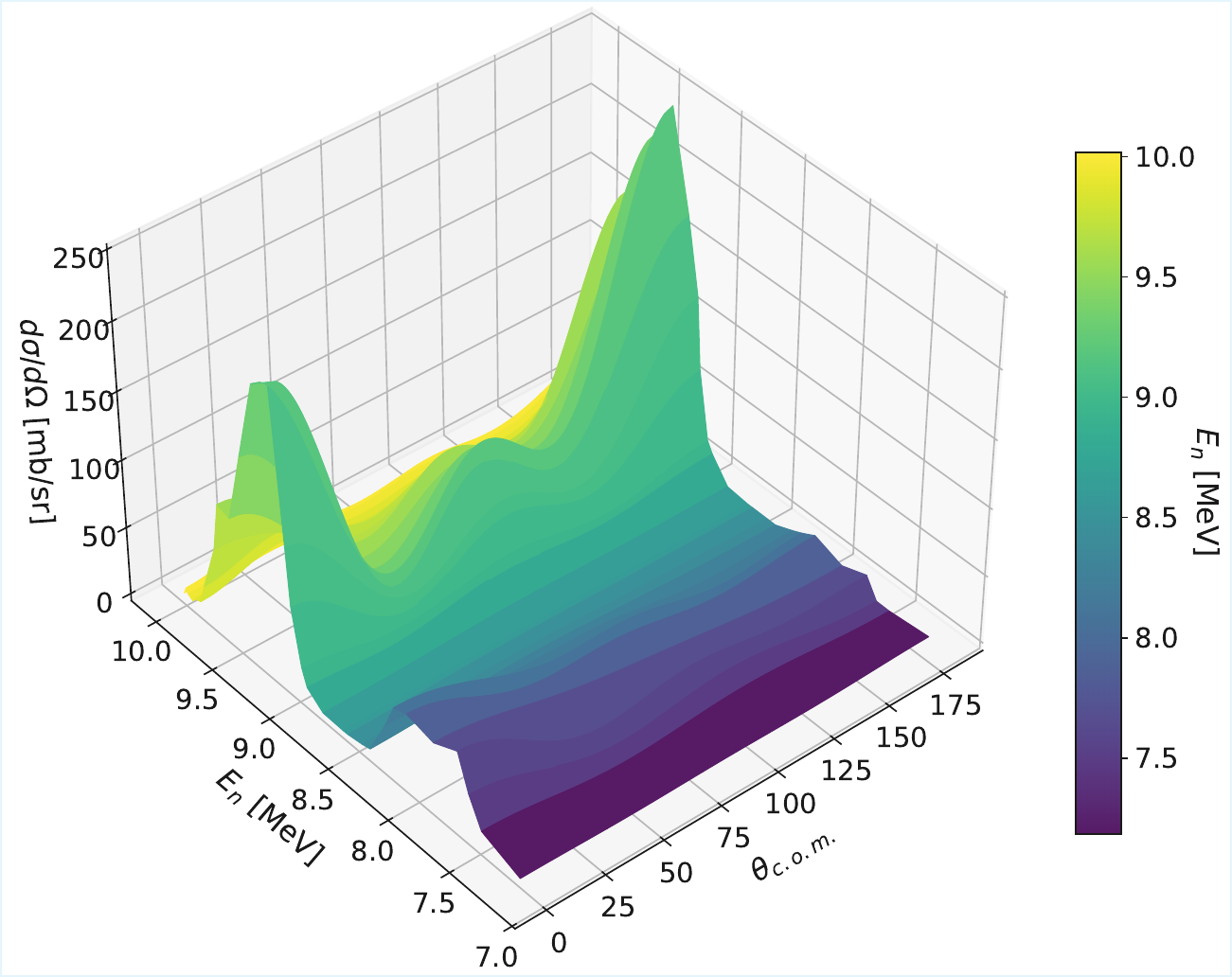}}
    \caption{Centre of mass differential cross sections from the $^{9}$Be($\alpha$,n$_0$)$^{12}$C reaction from Geiger \cite{Geiger,Geiger2}, converted to different incident neutron energies for the \c12 interaction. The colour scale also represents neutron energy to aid visualisation. }
    \label{fig:Geiger}
\end{figure}

The \c12 differential cross sections for the full selection of neutron energies are shown in Fig.~\ref{fig:diff12C}.
A fairly robust data set is that of the time-reversed $^{9}$Be($\alpha,n$) of Geiger and van Der Zwan \cite{Geiger}, originating from the experimental results of van Der Zwan and Obst \cite{VDZ,Obst} where the Legendre coefficients, $A_i$, up to order 8 were determined for a broad range of alpha-particle energies \cite{Geiger2}. The trend of the differential cross section with energy from the Geiger work is demonstrated in Fig~\ref{fig:Geiger}. The Geiger values, as shown by the red lines in Fig.~\ref{fig:diff12C} have been overlaid with our results where the absolute normalisation was taken by scaling the $A_0$ to be $\sigma_{n,\alpha}/(4\pi)$ from the work of Kuvin \cite{12CnaLANSCE}.\\
At the lowest neutron energies of $E_n$ = 7.25 MeV the agreement with the Geiger is poor, mainly due to the absolute magnitude of the cross section. In the Geiger data, the closest energy point for comparison is 75 keV lower in energy, and with a relatively large spread in neutron energy, the average \c12 cross section will be much larger than the normalisation value used here. As the energy increases, despite the smaller angular range covered at the higher gas pressure of 100 Torr, the results for $E_n \le 8.15$ MeV agree relatively well. At $E_n \ge 8.36$ MeV, the lower gas pressure of 50 Torr means a larger angular range can be covered and much closer agreement is found. This demonstrates the channel separation for the \c12 reaction has been achieved cleanly using the angular and length cuts discussed above. Additionally, the normalisation of our data from the $^{1}$H$(n,p)$ is also demonstrated to show consistent \c12 cross sections with previous well understood values \cite{12CnaLANSCE}. This point becomes particularly important for the \o16 cross section where historical differences have been observed.

\subsection{$^{16}\mathrm{O}(n,\alpha_0)$}
Obtaining $^{16}$O+$n$ cross sections are experimentally much more challenging than for $^{12}$C+$n$. This is due to the higher density of levels, with a large fraction being narrow.  
The total cross section for \o16 at lower energies has been influenced by historical confusion where different data sets saw a 30\% discrepancy for the time reverse {$^{13}\mathrm{C}(\alpha,n)$} reaction \cite{Sekharan,Harissopulos}. Due to the difficulties in understanding the neutron detection efficiency for this reaction, detector developments of neutron long counters with a constant efficiency with neutron energy have since been developed. This rectified the discrepancy as arising due to an overly simplistic model of the $(\alpha,n_i)$ components across different alpha-particle energies \cite{HeBGB1,HeBGB2}. As the incident energy changes with traditional neutron detectors, one becomes very sensitive in the total measured cross section to different populations of the $n_0$, $n_1$, $n_2$, etc. due to changes in the outgoing neutron energy and, therefore, the efficiency. The measurements of the time-reversed reaction conducted in this way cannot separate the $^{13}$C$(\alpha,n_0)$ channel of interest here. A full ENDF evaluation based on an R-matrix analysis of $^{17}$O \cite{ENDFVIII} covers neutron energies below 7 MeV using data from multiple channels. Cross sections above this range were `joined smoothly' between 6.5-7.5 MeV. For the \o16 reaction, there is still uncertainty about the total cross section above $E_n$ = 7 MeV and measurement of the total cross section (requiring full angular coverage of the recoil $\alpha$) is extremely challenging. Differential cross sections covering as broad an angular range as possible allow for the best possible angle-integrated cross sections with minimal reliance on predictions of the angular distribution outside the range covered by data.\par

\begin{figure}
    \centering{\includegraphics[width=\textwidth]{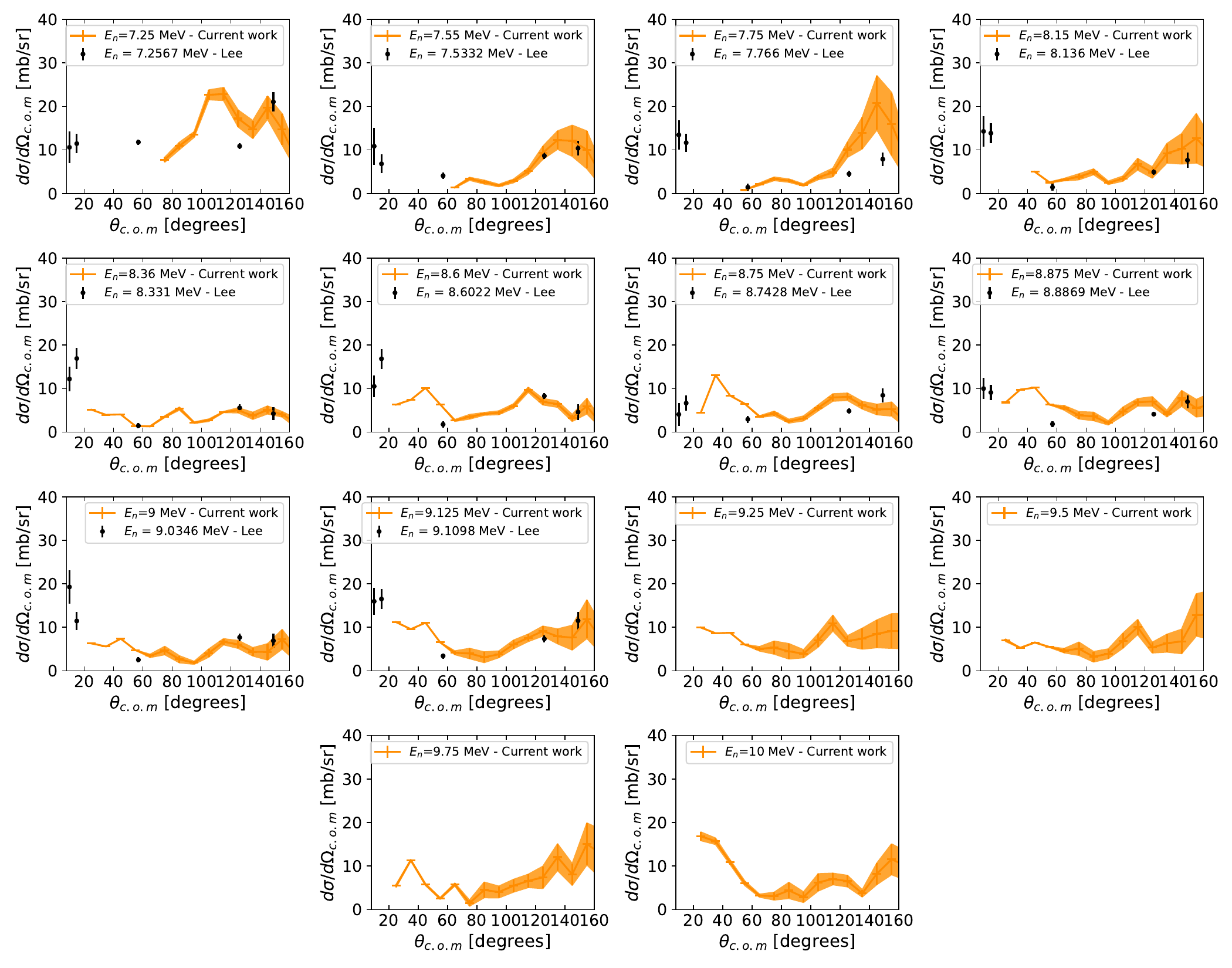}}
    \caption{Centre of mass differential cross sections for the \o16 reaction at different incident neutron energies in comparison to the values of Lee \cite{Lee}.}
    \label{fig:diff16O}
\end{figure}

Figure~\ref{fig:diff16O} shows the values for the \o16 differential cross section from the current work. At lower neutron energies, the cross section appears to peak at backwards angles. One may compare with the recent results of Lee \cite{Lee} using LENZ at LANSCE which provide high-quality \o16 differential cross section measurements over our entire energy range, albeit at a limited number of angles. There is a reasonable agreement across many different neutron energies at large centre of mass angles although our cross sections appear higher than those of Lee at $\theta_{c.o.m} = 57^{\circ}$ for neutron energies exceeding 8.6 MeV. The Lee differential cross section at this angle also showed disagreement from the time-reversed $(\alpha,n_0$) of Robb \cite{Robb} at $E_n$ = 7.23 MeV where the Lee value appears almost a factor of two larger. As in the current data, the neutron energy FWHM of the Lee data was rather large at around 0.13 MeV for $E_n$ = 7.25 MeV so complicated energy smearing effects near the strong resonance in $^{16}$O+$n$ at $E_n$ = 7.2 MeV may contribute to discrepancies between different measurements. In the current work for example, the total width ranges from 350-250 keV from $E_n$ = 7.2 - 10 MeV accordingly, with a smeared flat-top distribution \cite{JBNature}.\\

\subsection{$^{16}\mathrm{O}(n,\alpha_{1,2,3})$}
\begin{figure}
    \centering{\includegraphics[width=\textwidth]{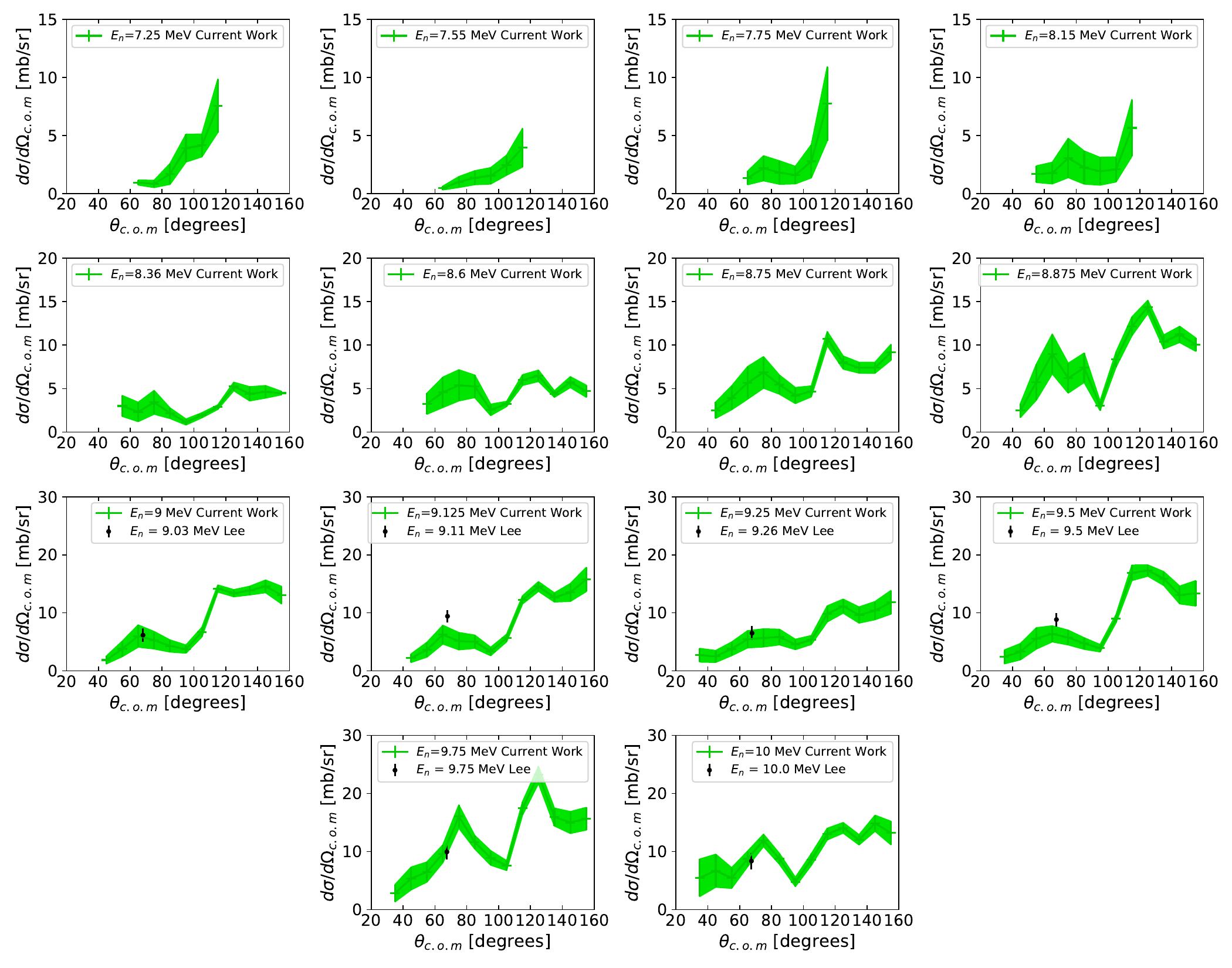}}
    \caption{Centre-of-mass differential cross sections for the \po16 reaction at different incident neutron energies in comparison to the values of Lee \cite{Lee}.}
    \label{fig:diff16Op}
\end{figure}

The \po16 cross section is difficult to obtain. As noted above, we cannot separate the $\alpha_{1,2,3,..4}$ components, so these are grouped together and data from the inverse $^{13}$C($\alpha,n$) reaction cannot be used. As with the \o16, previous LENZ results from Lee have measured this component \cite{Lee} (similarly grouped into $\alpha_{1,2,3}$) however with just one angle that overlaps the current data for $E_n > 9$ MeV. In this energy range, this component is expected to have a larger cross section than for directly populating the $^{13}$C ground state for $\alpha_0$. Our differential cross-section data for \po16 are shown in Fig.~\ref{fig:diff16Op} where a different behaviour from the \o16 can be seen with a regular minimum occurring around 90$^{\circ}$ suggesting odd Legendre polynomials contribute or the even polynomials destructively interfere. The comparison between the current work and a previously measured data point at $\theta_{c.o.m} \approx 68^{\circ}$ \cite{Lee} again shows reasonable agreement.\par
\section{Angle integrated cross section}
As expected from the ENDFVIII.0 \cite{ENDFVIII} evaluations due to the fact there are multiple channels, our data show that the \po16 differential cross section is larger than the \o16 above 9 MeV. Using our limited angular range, it is possible to estimate the total cross section for these two components. It is important to note that without information across the full angular range, the results here are only an estimator of the total cross section. Given that:
\begin{eqnarray}
\sigma = \int \frac{d\sigma}{d\Omega} d\Omega = 2\pi \int \frac{d \sigma}{d \Omega} \sin{\theta} d\theta,
\end{eqnarray}
with our larger angular coverage than Lee, rather than approximate a constant d$\sigma$/d$\Omega$, we opt to use a $\sin(\theta)$ weighting to obtain a better cross-section estimate. Given our data do not cover $\theta$ from $[0,\pi]$, one must carefully account for these unmeasured regions and can assign a more constrained estimate of the constant value for these angles. Given the sine weighting, these larger angles however are of smaller importance so the model error correspondingly becomes smaller. To account for this, our estimate of the total cross section as the angle-integrated cross section, $\tilde \sigma$, is calculated by the angular weighting:
\begin{equation}
\tilde\sigma =  2\pi\frac{\left (\int_{\theta_l}^{\theta_h} \frac{d \sigma}{d \Omega} \sin{\theta} d\theta \right) \left (\int_{0}^{\pi} \sin{\theta} d\theta \right)}{\left (\int_{\theta_l}^{\theta_h} \sin{\theta} d\theta \right)} =  4\pi\frac{\int_{\theta_l}^{\theta_h} \frac{d \sigma}{d \Omega} \sin{\theta} d\theta }{\int_{\theta_l}^{\theta_h} \sin{\theta} d\theta },\label{eq:sigmatilde}
\end{equation}
to account for the limited angular range between $\theta_L$ and $\theta_H$. It is worth noting that in cases of isolated resonances, the even Legendre coefficients are often positive and this correspondingly creates a maximal contribution at forward and backward angles. As such, this method is likely to systematically underestimate the cross section. In the excitation regime of multiple resonances, this effect becomes reduced and the angle-integrated cross section becomes less likely to underestimate the total cross section. One may compare this method against obtaining the cross section via Legendre polynomials by a fit up to arbitrary order. This technique is seen to be more susceptible to poor estimation of the error on the cross section obtained by the coefficient for the zeroth order Legendre polynomial, $A_0$, due to strong correlations between even Legendre polynomials when the fit does not cover a significant fraction of $\theta$ from $[0,\pi]$.

Our incident neutron energies are not normally distributed, arising from the dominant contribution of energy loss through the 7.98-cm-long deuterium gas cell sweeping out many neutron energies, and energy straggling. Therefore, the energy error bars used represent the total width of the neutron energy spectrum. The cross section error bars correspond to contributions from the integration of errors from the $(n,\alpha)$ differential cross section in combination with counting statistics from the normalisation $^{1}$H$(n,p)$ and do not incorporate model errors due to limited angular range information.\par
\begin{figure}
    \centering{\includegraphics[width=0.6\textwidth]{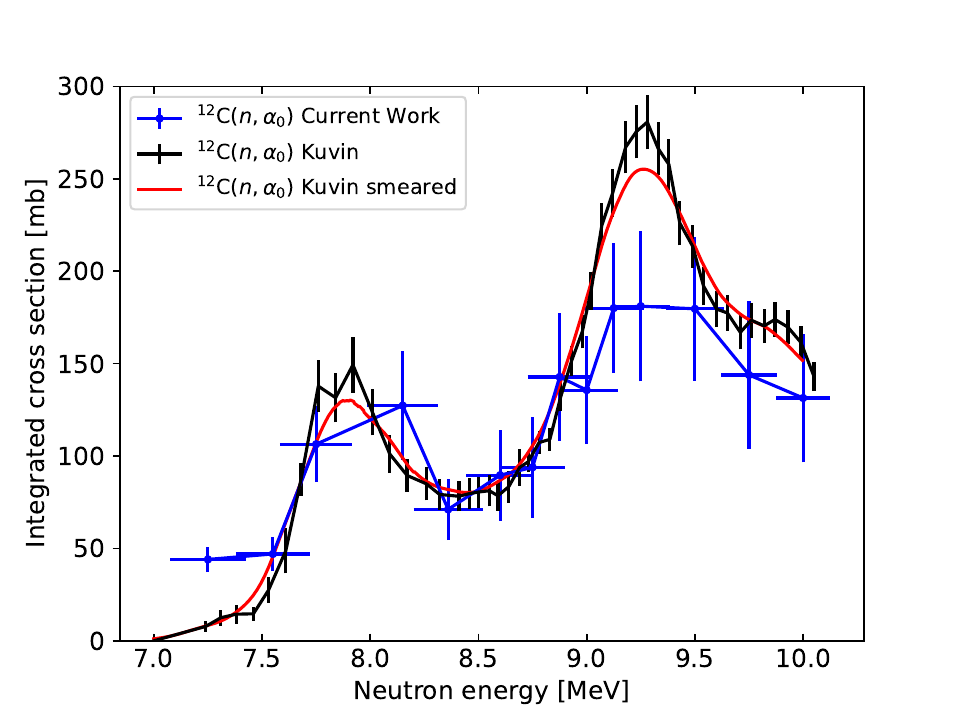}}
    \caption{Angle-integrated cross section, $\tilde \sigma$, for \c12 across the angular range covered by the current work, compared to previous literature values of Kuvin \cite{12CnaLANSCE} which has also been smeared by our intrinsic beam energy spread, shown by the red line.}
    \label{fig:total12C}
\end{figure}
\begin{figure*}
    \centering{\includegraphics[width=0.8\textwidth]{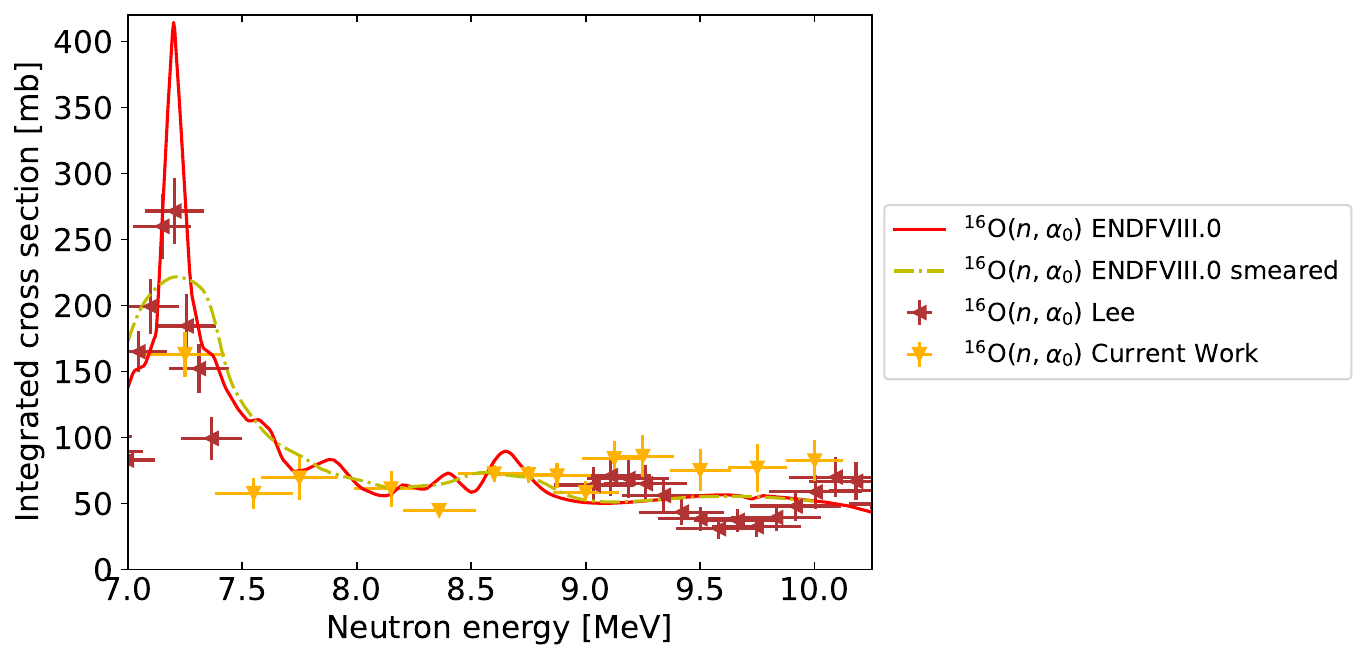}}
    \centering{\includegraphics[width=0.8\textwidth]{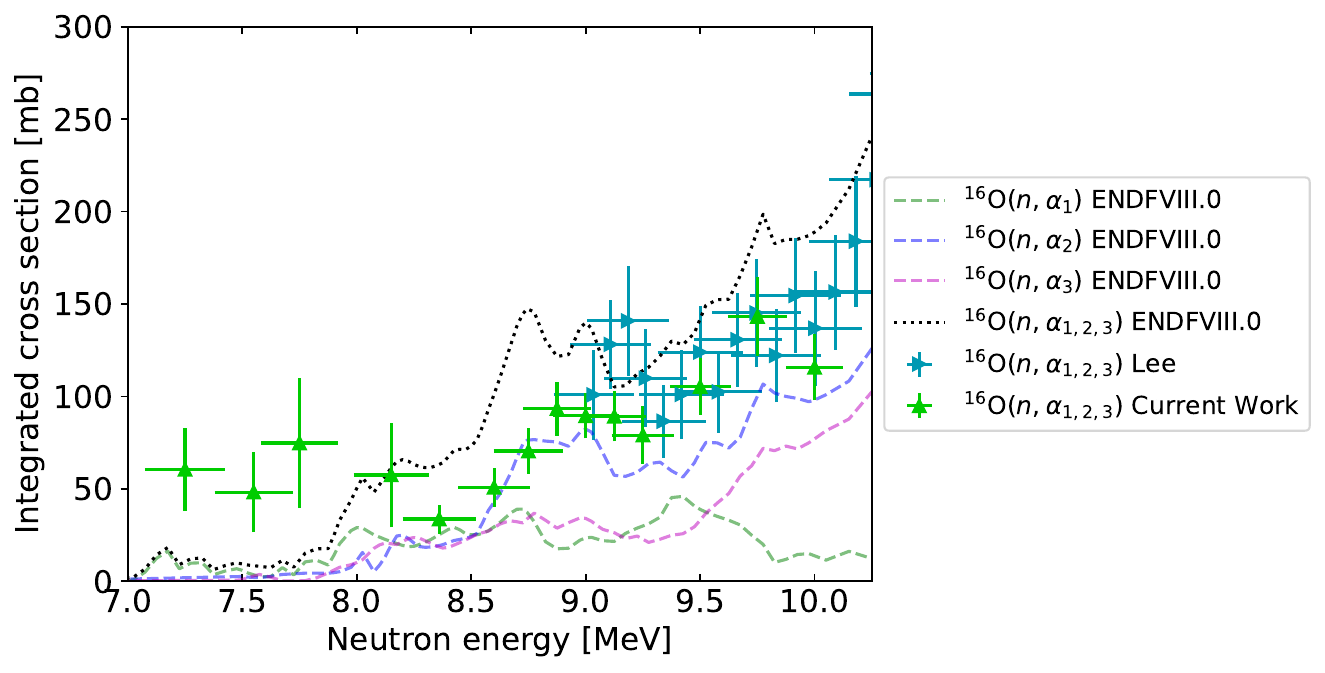}}
    \caption{Angle-integrated cross section, $\tilde \sigma$, for \o16 (left) and \po16 (right) across the angular range covered by the current work, compared to previous literature values of Lee \cite{Lee} and the ENDFVIII.0 evaluation \cite{ENDFVIII}. The \o16 ENDFVIII.0 cross section is also shown after being smeared with our intrinsic beam energy spectrum.}
    \label{fig:total16O}
\end{figure*}

To compare our data against the evaluation libraries, on Figs.~\ref{fig:total12C} and \ref{fig:total16O} the smeared ENDFVIII.0 cross sections are plotted. Here, the cross section is convoluted with our neutron energy spectrum (taking into account changes in width over the total neutron energy range) which should give a comparable cross section value when at the centroid of our data points.\par
The values obtained from this method for \c12 are shown in Fig.~\ref{fig:total12C} in comparison with the results of Kuvin \cite{12CnaLANSCE}. Reasonable agreement is observed over much of the energy range, only with an over-estimated cross section at small neutron energies, as discussed above due to the rapidly rising cross section and broad energy profile of our beam. The angle-integrated cross section of the current work around the peak at $E_n$~=~9.2 MeV is also smaller than the previously measured results. This discrepancy may be attributed to large cross sections at both small and large angles that are outside of our covered angular range and highlight the systematic underestimate possible for strong isolated resonance discussed above, meaning that our angle-integrated cross section differs from the more reliable total cross section results previously obtained.

The angle integrated cross section for \o16 and \po16 are shown in Fig.~\ref{fig:total16O}. They are shown in comparison to the angle-integrated cross section of Lee (where they make the assumption that the differential cross section is the same for all angles as the average of the $\theta = 15^{\circ}$ and $\theta = 57^{\circ}$ values) and the ENDFVIII.0 evaluation \cite{ENDFVIII}. The individual ENDF components for $(n,\alpha_1)$, $(n,\alpha_2)$, and $(n,\alpha_3)$ are also shown separately.\par
The \o16 values show good agreement (despite limited angular range) with the ENDFVIII evaluations (MT=800) at the smaller neutron energies with the value near the resonance at $E_n$ = 7.25 MeV being consistent in the current work. Our values match well those of Lee at this energy. As the neutron energy exceeds 7.5 MeV, our \o16 cross section also follows well that of the ENDF values. For $E_n > 9$ MeV, our values can once again be compared to those of Lee where the current work shows a much flatter energy behaviour, similar to the ENDF values. The relative uncertainty of our \o16 integrated cross sections range from 7-24\% across the different energy bins. This exceeds the desired 5\% accuracy across the 2-20 MeV neutron energy range \cite{HPRL,HPRLweb}, but is a significant improvement on the previous 30\% uncertainty. Once propagated through nuclear data libraries, with a 30\% uncertainty in $^{16}\mathrm{O}(n,\alpha)$ leading to a 100 pcm (i.e. $10^{-3}$ relative) uncertainty for $k_{eff}$, these new results can hopefully contribute to rectifying an under-prediction of light water reactors' reactivity \cite{NEA}.\par
The \po16 cross section components show a larger than expected integrated cross section at lower energies with the value of around 50 mb from the current work exceeding the ENDF value (MT=801, 802, 803)  by a factor of 5. As the energy increases however, the ENDF results begin to exceed our integrated cross section, which have good agreement with the Lee results despite limited angular range. The current results show the importance of measuring the differential cross sections for as broad a range of angles as possible which can have significant impact on the evaluated libraries' total cross sections. The relative uncertainties for this channel range from 45\% for $E_n < 8.36$ MeV, to around 15\% for $E_n \ge 8.36$ MeV. While the $^{16}\mathrm{O}(n,\alpha)$ reaction now appears to be well understood below $E_n$ = 7 MeV, there is clearly a great need for extending the R-Matrix evaluation to higher energies, which is a much better approach for determining angle-integrated cross sections from differential data over a limited angular range. The considerable contribution of the \po16 to the total $^{16}\mathrm{O}(n,\alpha)$ cross section means greater neutron loss inside of reactors, modifying $k_{eff}$, and causing greater He build-up than previously accounted for, potentially precipitating helium embrittlement effects in oxygen-containing reactor facing materials.

\section{Discussion}
This work demonstrates that neutron-induced reactions can be measured with active-target TPCs. Current limitations around track-length reconstruction and angular acceptance can be rectified by future detector improvements such as increasing the height of the field cage to minimise escaping light recoils in the vertical direction which will provide an additional route for channel selection. Having a larger active area coupled with a decreased gas pressure, or increased pad granularity will also improve the charged-particle track quality and therefore provide better angular resolution. Currently, the angular binning is dictated by statistics rather than resolution, but improvements in the incident neutron flux (such as primary gas target development to facilitate an increased primary beam current) may generate significant statistics in short runs.\par
Running in active-target mode, there are obviously limits on which reactions can be measured as it must be a component of a gas with good TPC properties or coupled with a quenching gas with a well understood background contribution. Incorporating a gas with a hydrogen component also incorporates a built-in normalisation method through the $^{1}$H$(n,p)$ reaction, as used in the current work, but without the requirement for a CH$_2$ conversion foil.
\section{Conclusion}
The \c12, \o16 and \po16 differential cross sections were obtained across a variety of neutron energies using the TexAT TPC. Channel separation was made by distinguishing between the angle of the heavy and light recoils, and the range of the heavy recoil in the gas. A multinomial model was used to propagate ambiguities in channel selection on an event-by-event basis. The results for the three reaction channels appear to agree well with previous data where overlapping angles and energies are available. For the \o16 and \po16, our current data provide useful differential cross section values across a much broader angular range than previously measured. This will be of great use for expanding the existing R-Matrix description of $^{17}$O up to $E_n$ = 10 MeV to describe the $^{16}$O($n,\alpha$) in greater detail. This reaction is of interest in modelling the neutron multiplication factor, $k_{eff}$ in nuclear reactors and our results suggest the evaluation libraries have a poor understanding of the cross section above $E_n$ = 7.5 MeV.\par
This work demonstrates that future nuclear data experiments are possible using a Time Projection Chamber to study neutron-induced reactions.

%
%


\funding{This work was supported by the U.S. Department of Energy, Office of Science, Office of Nuclear Science (under awards no. DE-FG02-93ER40773 (J.B., G.V.R., S.A., E.K.), DE-FG02-88ER40387 (K.B., C.R.B., G.H., J.A., T.N.M., Z.M., S.P., M.S., N.S., D.S., S.S., A.V.V., J.W.), DE-SC0019042 (Z.M., M.S., N.S., D.S.), DE-NA0003883 (C.R.B., Y.J.A., T.N.M., S.P., J.W.), DE-NA0003909 (K.B., Z.M., A.V.V.), DE-FG02-87ER-40316 (R.J.C., N.D. and L.G.S.), by National Nuclear Security Administration through the Center for Excellence in Nuclear Training and University-Based Research (CENTAUR) under grant number DE-NA0003841 (J.B., C.E.P., G.V.R., S.A., E.K., N.D., E.V.O., L.G.S.) and by the UK STFC Network+ Award Grant number ST/N00244X/1 (Tz.K., R.S., and C.W.) and STFC Grant number ST/Y00034X/1 (Tz.K. and C.W) and ST/Y000331/1 (R.S.). This work also benefited from support by the National Science Foundation under Grant No. PHY-1430152 (JINA Center for the Evolution of the Elements). G.V.R. also acknowledges the support of the Nuclear Solutions Institute.\\
For the purpose of open access, the author has applied a Creative Commons Attribution (CC BY) license to any Author Accepted Manuscript version of this paper arising from this submission.}


\data{The cross section data contained within this paper are available at: https://doi.org/10.25500/00001468.}


\bibliography{references}

\end{document}